\documentclass[journal=jacsat,manuscript=article]{achemso}
\setkeys{acs}{keywords = true}

 \pdfoutput=1

\usepackage{chemformula} 
\usepackage[T1]{fontenc} 

\usepackage{graphicx}
\usepackage{color}
\usepackage{epsfig}
\usepackage{subfigure}
\usepackage{natbib}
\usepackage{array}
\usepackage{amsmath}
\usepackage{ulem}
\usepackage{mathtools}
\usepackage{hyperref}
\hypersetup{pdfnewwindow=true, colorlinks=true, linkcolor=blue, anchorcolor=blue,citecolor=blue, filecolor=blue,menucolor=blue, urlcolor=blue}

\DeclarePairedDelimiterX\braket[2]{\langle}{\rangle}{#1\,\delimsize\vert\,\mathopen{}#2}
\usepackage{mhchem}

\usepackage[table]{xcolor} 

\setkeys{acs}{articletitle=true}
\setkeys{acs}{maxauthors = 10, etalmode=truncate}      
\makeatletter
\newcommand*{\forcekeywords}{
  \acs@keywords@print
  \let\acs@keywords@print\relax
}
\makeatother

\author{Suryakanti Debata}
\affiliation{Laboratory for Physical Sciences, College Park, Maryland 20740, USA}

\author{Sai Krishna Narayanan}
\affiliation{Department of Physics, University of Maryland, College Park, Maryland 20742, USA}

\author{Pratibha Dev}
\affiliation{Laboratory for Physical Sciences, College Park, Maryland 20740, USA}

\email{pdev@lps.umd.edu}
\date{\today}

\title{H\MakeLowercase{idden} \MakeLowercase{in} P\MakeLowercase{lain} S\MakeLowercase{ight}: A\MakeLowercase{romaticity} of H\MakeLowercase{exagonal} B\MakeLowercase{oron} N\MakeLowercase{itride}}

\begin{document}

\begin{abstract}
Hexagonal boron nitride (hBN) and graphene are similar in many ways - they are isoelectronic, have the same structure, are chemically inert and show persistence. All of these properties are indicators of a deeper connection that has, thus far, been overlooked. Unlike graphene, which has been shown to be aromatic, it is not known if hBN is aromatic.  In this density functional theory-based work, we investigate the aromaticity (or lack thereof ) of hBN.  By employing the magnetic criterion, supported by group theoretic and energetic considerations, we show that hexagonal boron nitride is indeed aromatic, even if weakly so, as compared to graphene.   Since 
aromaticity is used to understand physical and chemical properties of planar compounds, the picture developed in this work is important to bridging the gap between the physical and chemical understanding of hBN's properties.

\end{abstract}

\maketitle


\section*{Introduction}

 Hexagonal boron nitride (hBN) is a two-dimensional (2D)  inorganic structural analogue of graphene, with alternating $sp^{2}$-hybridized boron and nitrogen atoms arranged in a 2D honeycomb lattice.   Nicknamed ``white graphene'', hBN  is a wide band gap semiconductor with a gap of 5.96\,eV~\cite{Cassabois_indirectGap_2016}. Its insulating nature, atomically-flat structure, chemical and mechanical stability have made hBN essential for a range of applications, as either an active or a passive component~\cite{Dean2010,Lee_hBN_Substrate_2013,IgorAharonovich2015,Cao2018,Dehghani2018_hBN_biosensing,Fang_hBN_Substrate_2020,Saikrishna2023,Fukamachi2023ER,Stern_2024_hBN_spinmemory,Lee_2025_hBN_spinmemory}.  For such a simple and technologically-important material, a fundamental question remains unanswered: ``Is hBN aromatic?''

 Aromaticity is a powerful and useful concept in chemistry. It is used to rationalize the structure, reactivity and persistence of cyclic/polycyclic materials.  Historically, the concept of aromaticity, introduced by August Kekul\'{e}~\cite{Kekule_1865_aromaticity1, Kekule_1865_aromaticity2}, was used to explain the stability of planar molecules, such as benzene and polycyclic hydrocarbons, which exhibited cyclic delocalization of $\pi$-electrons and a resonance between different possible forms for each structure. In recent years, however, the concept of aromaticity has evolved (for better or worse~\cite{Hoffmann2015}). For example, it is now invoked for a broader range of structures, such as non-planar molecules~\cite{Antic_2017_aromatic_nonplanar}, molecular cages~\cite{Li_2005_aromatic_cage}, inorganic molecules~\cite{Islas2007}, and even excited-state species~\cite{Baird_1972_aromatic_excited, Baranac_2020_aromatic_excited}.  The 2004 discovery of graphene~\cite{Novoselov2004} resulted in yet another expansion of the concept of aromaticity, this time to an infinite 2D crystal~\cite{Popov2012,Zdetsis_graphene2016,Zdetsis_graphene2020}.  In our work, we explore the aromaticity of hBN, making connections between the physical and chemical properties of this 2D crystal.

There are several reasons why aromaticity of hBN has not been previously explored. One of the more obvious reasons is that as a concept, aromaticity mostly concerns chemists, while as a crystal, hBN and its properties are more often of interest to physicists. So even though chemists have been debating for several decades about the aromaticity (or lack thereof) of borazine [chemical formula: $\mathrm{B_{3}N_{3}H_{6}}$], which can be regarded as a smallest unit/flake of hBN~\cite{Schleyer1997_borazine,Benker_2005_Borazine,Islas2007,baez2022}, a similar consideration has not been given to hBN.  On the other hand, it may even seem that the matter is already settled and hBN is aromatic, just like graphene~\cite{Popov2012}. This is owing to the fact that it already satisfies the following criteria for aromaticity: (i) H\"{u}ckel's $4n+2$ $\pi$-electrons rule~\cite{Huckel1931} per ring ($n$ is a positive integer), (ii) planarity of structure, (iii) equal bond lengths, and (iv) enhanced stability.  However, most of the aforementioned aromaticity criteria satisfied by hBN have several known exceptions in chemistry~\cite{Hoffmann2015,Schleyer_Review_2005}, and by themselves these criteria do not represent definitive proof of aromaticity. In addition, in spite of being structurally similar and isoelectronic to graphene, which is aromatic~\cite{Popov2012}, hBN's electronic structure properties are quite unlike those of graphene. This is due to the difference between the electronegativities of boron (2.04) and nitogen (3.04), which makes hBN a mixed ionic-covalent compound with polar bonds. Due to the higher electronegativity of N-atoms, the $\mathrm{sp^{2}}$ hybridized $\sigma$-electrons are localized closer to the nitrogens. The boron atoms have empty $2\mathrm{p_{z}}$-orbitals, while the $2\mathrm{p_{z}}$-orbitals of the nitrogens contain a lone pair. It should be mentioned that the $2\mathrm{p_{z}}$-orbitals of boron atoms are only formally empty -- the lone pairs of nitrogens are partially delocalized towards the borons~\cite{Izyumskaya_2017_hbn, Weng_2016_hbn, Gong_2021_hbn}.  These differences in electronic structure relative to graphene, makes it important to conduct a careful investigation of the $\pi$-electron delocalization and its consequences for hBN.

In the present density functional theory (DFT)-based work, we explored hBN's aromaticity by creating finite flakes (molecular models) of hBN, as was done previously by works on graphene's aromaticity~\cite{Popov2012,Zdetsis_graphene2016,Zdetsis_graphene2020}.  This allowed us to employ different chemical measures/criteria, which were primarily developed for finite structures, to investigate hBN.  For sake of establishing the accuracy of our analysis and for comparison, the aromaticity of graphene flakes was also studied.  The magnetic criterion,  which is based on the response of the delocalized $\pi$-electrons to an applied magnetic field, is considered to be the most reliable determinant of aromaticity. Hence, in our work, we used it to classify and quantify the aromaticity of hBN.  This analysis is supported by the energetic criterion and group theoretic considerations that help to develop a fuller picture of hBN's aromaticity.

  \section*{Results and Discussion}

\begin{figure*}[t]
    \centering
    \includegraphics[width=0.98 \linewidth]{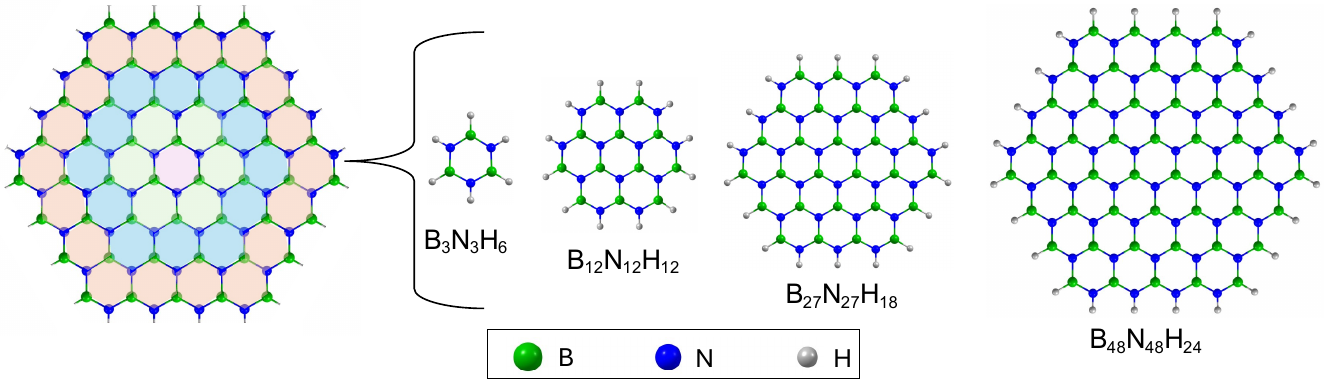}
    \caption{ Flakes with the general chemical formula of B$_{3n^{2}}$N$_{3n^{2}}$H$_{6n}$, $n=1, 2, 3,...$ (shown on the right) created from a bulk hBN monolayer (on the left).   The smallest possible flake, borazine ($n=1$), is obtained from the hydrogenated pink hexagonal region of 2D hBN. Larger flakes are obtained by including additional hBN rings, which are indicated by the different colored regions of the 2D hBN on the left. }
    \label{fig:Fig1}
\end{figure*}

 Our study utilized hexagonal flakes of hBN and graphene of different sizes. Starting with borazine (B$_{3}$N$_{3}$H$_{6}$) and benzene (C$_{6}$H$_{6}$), which can be treated as basic units of hBN and graphene, respectively, we created larger polycyclic hexagonal flakes by increasing the number of rings at the periphery, creating a sequence of flakes as seen in Figure~\ref{fig:Fig1}. To saturate the peripheral dangling bonds, we passivated the model systems with hydrogen atoms, ensuring that all atoms are fully coordinated, giving hBN (graphene) flakes with the general formula of B$_{3n^{2}}$N$_{3n^{2}}$H$_{6n}$ (C$_{6n^{2}}$H$_{6n}$), $n=1, 2, 3,...$.  In what follows, we use different criteria to determine the aromaticity of hBN and compare it with the aromatic properties of graphene.

 \begin{figure*}[t]
    \centering
    \includegraphics[width=0.95 \linewidth]{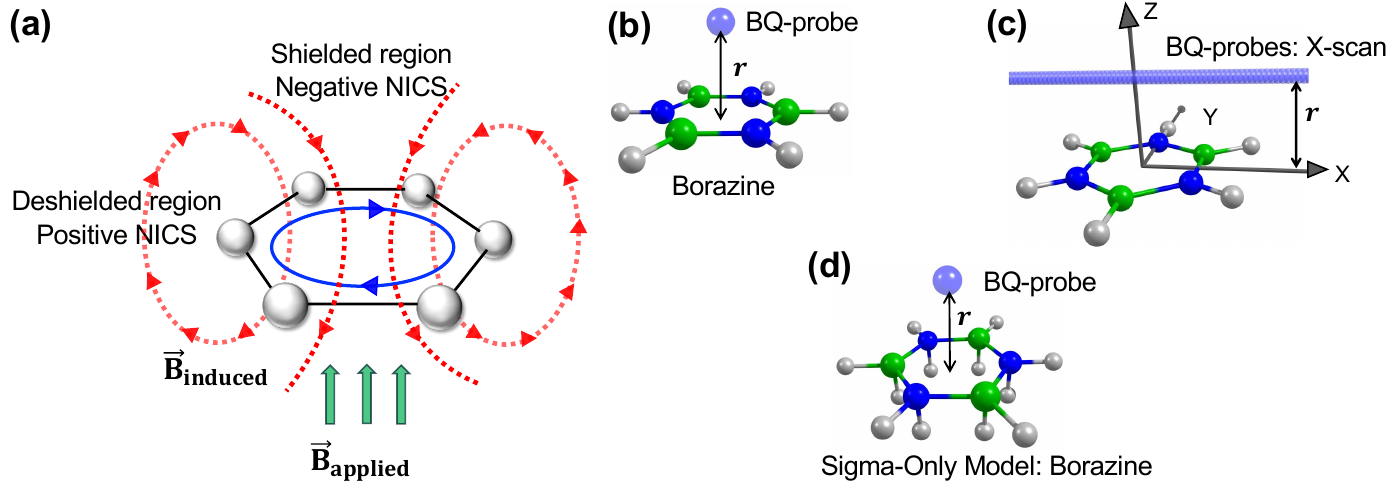}
    \caption{ Magnetic criterion for aromaticity: (a) Magnetically-induced aromatic ring current (in blue) and the resulting shielding (deshielding)
effect of the current-induced magnetic field (in red) in the endocyclic (exocyclic) region. (b) The BQ-probe (ghost atom), shown in purple, used to obtain the NICS-index at a distance $r$ above the ring center. (c) A series of BQ-probes placed along the X-direction to survey the NICS-indices across the molecule. (d) The $\sigma$-Only model for borazine obtained by placing hydrogen atoms below the B- and N-atoms. }
    \label{fig:Fig2}
\end{figure*}

\subsection{Magnetic Criterion of Aromaticity}

Schleyer's nucleus independent chemical shift (NICS) method~\cite{Schleyer_1996_NICS,Schleyer_1997_NICS1} is a straightforward and reliable criterion for evaluating aromaticity or antiaromaticity by measuring the response of a ringed structure to an externally applied magnetic field~\cite{Chen2005_NICS,Stanger2006,Stanger_2010_NICS_SOM,Gershoni2015,Stanger2020}. In response to the external magnetic field, the delocalized $\pi$-electrons in a monocyclic or polycyclic structure generate an induced ring current (shown in blue in Figure~\ref{fig:Fig2}a), which in turn creates its own magnetic field (shown in red in Figure~\ref{fig:Fig2}a).  The magnetic shielding due to the induced ring current can be obtained at a point of interest by placing a ghost atom (labelled as the ``BQ" probe) at that point, as illustrated in Figure~\ref{fig:Fig2}b. Since the ghost atom has no protons or electrons, it is a passive observer of the magnetic shielding. 
The calculated magnetic shielding (with the sign reversed) gives the NICS value at that point.  Aromatic rings display diatropic (clockwise) ring currents, which produce negative (shielded) NICS values. On the other hand, positive (deshielded) NICS values reflect paratropic (i.e., anticlockwise) ring currents, signifying antiaromatic behavior.  There are different possible constructs for obtaining NICS indices, which can broadly be classified as: (i) single-point NICS methods and (ii) multi-point NICS approaches.  Our work uses a combination of both classes of methods in order to create a fuller picture of hBN's aromaticity.  

 All single-point NICS approaches use a single BQ-probe (per ring) for aromaticity determination. 
A standard practice in the field is to place a BQ-probe at a certain distance, $r$, above the ring center (see Figure~\ref{fig:Fig2}b), thereby reducing the effects of $\sigma$ electrons and yielding NICS($r$)-values~\cite{Schleyer_1997_NICS1}.  
Usually, the probe height of $r=$1.0\,\AA{} is chosen to determine the isotropic NICS(1) index from (the negative of) the isotropic chemical shift at that point.   However, neither the choice of the isotropic chemical shift, nor the distance of  $r=$1.0\,\AA{} are necessarily appropriate for aromaticity determination. Since one is ultimately interested in the magnetic shielding due to the induced $\pi$-current, a better indicator of $\pi$-electron delocalization (as compared to the isotropic NICS value) is the out-of-plane component of the NICS tensor, i.e., NICS(r)$_{zz}$ (assuming the molecular plane to be the XY-plane)~\cite{Fowler1999electric, Steiner2001counter}.  This is the tensor component that we report in our work.  Moreover, since a single NICS value, computed at a fixed height above the ring's geometric center, may not adequately describe the spatial variation of diatropic and paratropic effects across extended, polycyclic hBN and graphene flakes, we performed multi-point NICS-XY-scan calculations across all flake models~\cite{Stanger2006,Stanger_2014_nicsscan}. In this approach, one uses multiple BQ probes across the two lateral directions, producing X- and Y-scans, and yielding the NICS profile for a molecule (see Figure~\ref{fig:Fig2}c).
 
 
Keeping in mind that the contributions from $\sigma$-electrons can still contaminate the NICS($r$)$_{zz}$ indices, one can adopt two approaches to avoid or minimize $\sigma$-effects: (i) use $\mathrm{NICS}_{\pi zz}$ methods, such as the $\sigma$-Only method (SOM), which are more effective in separating and removing contributions of $\sigma$-electrons to chemical shielding, but are more involved, or (ii) determine the optimal height at which $\sigma$-effects, which decay faster than $\pi$-effects, become negligibly small, and use this height to obtain the more-straightforward NICS($r$)$_{zz}$ indices (instead of using  $r=$1.0\,\AA{}). In our work, we adopted the latter approach using the procedure proposed by Gershoni-Poranne and  Stanger~\cite{Stanger_2014_nicsscan}. The process involves determining both $\mathrm{NICS}(r)_{\pi zz}$ and $\mathrm{NICS}(r)_{zz}$ values for a small subset of representative molecular models. In our case, we used borazine and benzene as stand-ins for hBN and graphene, respectively.  The $\mathrm{NICS}(r)_{\pi zz}$ values within the SOM approach~\cite{Stanger_2010_NICS_SOM} were calculated by adding hydrogen atoms at a fixed distance (1.0\,\AA{}) below each of the ring atoms that participate in the $\pi$-system, with the ghost-atom placed above the plane of the molecule as shown in Figure~\ref{fig:Fig2}d. Following this, the NICS values obtained for the hydrogenated fictitious system with $\sigma$-only contributions ($\mathrm{NICS}_{\sigma zz}$) are subtracted from those for the real system with contributions from both $\pi$ and $\sigma$ electrons to get $\mathrm{NICS}_{\pi zz}$, which should ideally have $\pi$ contributions only [i.e. $\mathrm{NICS}_{\pi zz}=\mathrm{NICS}_{zz}-\mathrm{NICS}_{\sigma zz}$]. The $\mathrm{NICS}(r)_{\pi zz}$ values obtained for different probe heights serve as a guide to determine the optimum height at which the $\sigma$ contributions in NICS(r)$_{zz}$ become negligible. This happens at an $r$-value at which the NICS-scan profiles of $\mathrm{NICS}_{\pi zz}$  and NICS$_{zz}$ match with each other.


\subsubsection{Optimal height determination for NICS($r$)$_{zz}$}

\begin{figure*}[t]
    \centering
    \includegraphics[width=0.95 \linewidth]{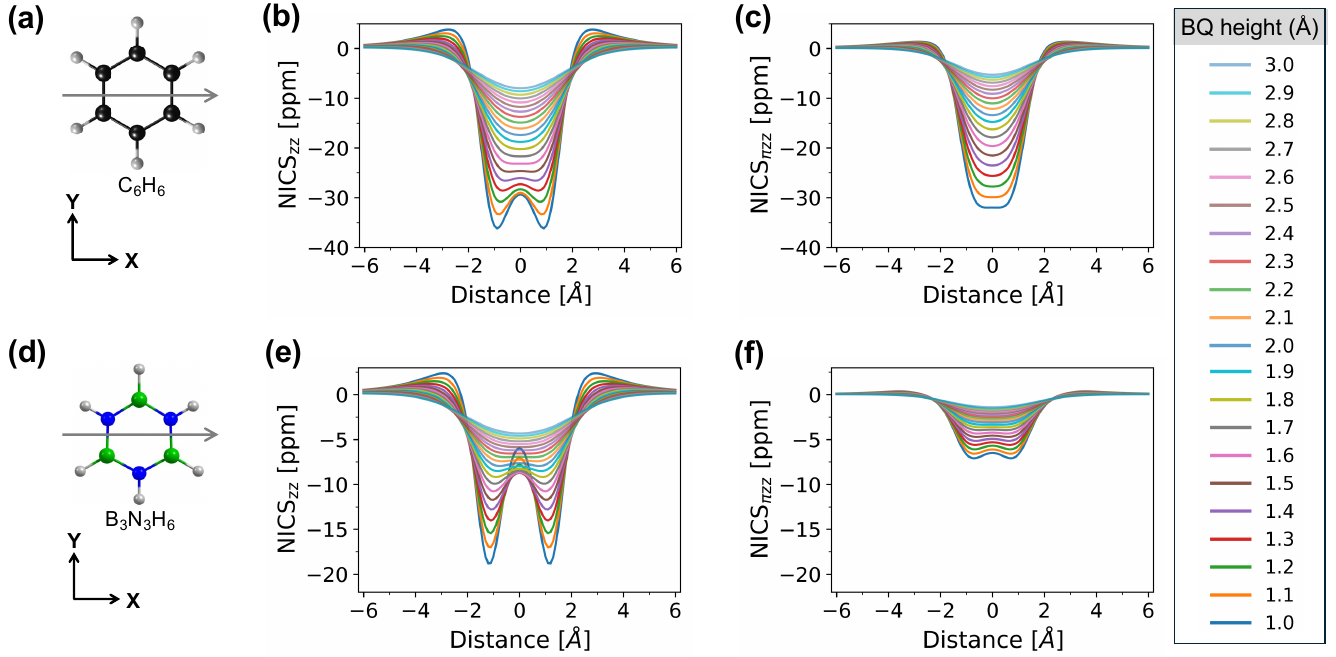}
    \caption{ (a) The lateral scan direction (X-axis) is shown by the gray arrows along which NICS values are calculated by placing multiple ghost atoms on benzene.  NICS-X-scans yielding (b) $\textrm{NICS}_{zz}$ and (c) $\textrm{NICS}_{\pi zz}$ values at different heights (1.0 - 3.0\,\AA {} in 0.1\,\AA {} increments) for benzene. In (b) and (c), the distance along the x-axis is given relative to the  geometrical center of the ring. (d) X-scan path (indicated by the gray arrow), (e)  $\textrm{NICS}_{zz}$ X-scan, and (f) $\textrm{NICS}_{\pi zz}$ scan values for borazine. Note that the y-axis scales used to plot NICS-scans of borazine are different from those for benzene.}
    \label{fig:Fig3}
\end{figure*}

To determine the appropriate height for the ghost atom at which the contribution to the magnetic shielding is mostly from $\pi$-electrons, $\mathrm{NICS}_{zz}$ calculations were performed with the BQ-probes at various heights $($1.0 - 3.0\,\AA {}, in 0.1\,\AA {} increments$)$ above the molecular planes of benzene and borazine. In Figures~\ref{fig:Fig3}a and d, the gray arrows indicate the X-axes scan path along which the in-plane/lateral scans were carried out. Figures~\ref{fig:Fig3}b and e show the $\mathrm{NICS}_{zz}$ values along the scan path for different heights above the molecular plane. 
Qualitatively, the $\mathrm{NICS}_{zz}$ scan profiles for benzene and borazine are quite similar, with both showing diatropicity (negative $\mathrm{NICS}_{zz}$ values) for an X-scan path within the molecules.  At lower heights, both borazine and benzene exhibit two prominent minima located at the bond centers rather than at the geometric centers of the rings. Such off-center minima indicate strong $\sigma$-contributions to chemical shielding~\cite{Stanger_2014_nicsscan}. As the $\sigma$-contribution decreases at larger heights of the BQ-probes, the minima move towards the ring center and finally merge to give one minimum at the ring center (also see Figure S1a in SI). In the case of borazine, the merging of the minima occurs at much higher probe heights as compared to benzene.  

Figures~\ref{fig:Fig3}c and f are plots of $\mathrm{NICS}_{\pi zz}$ scans using the more involved SOM approach for the two molecules.  The $\mathrm{NICS}_{\pi zz}$-scans for benzene in Figure~\ref{fig:Fig3}c show the expected behavior --  there is one minimum around the center of the ring, with the NICS-values changing rapidly as one approaches the circumference of the ring, outside which the expected paratropic effect can be seen.  For benzene, at  probe-heights of 1.7\,\AA{} and above, the profiles of NICS(r)$_{zz}$-scans in Figure~\ref{fig:Fig3}b becomes similar to those of  $\mathrm{NICS}_{\pi zz}$-scans in Figure~\ref{fig:Fig3}c. This implies that the appropriate height for the ghost atoms is 1.7\,\AA{} (and above) for benzene, in agreement with results of Gershoni-Poranne and Stanger~\cite{Stanger_2014_nicsscan}.  Unlike the single minimum in the $\mathrm{NICS}_{\pi zz}$ scans for benzene, $\mathrm{NICS}_{\pi zz}$ scans for borazine continue to show two minima at lower probe-heights (see Figure~\ref{fig:Fig3}f), even though the depths of the minima in the SOM NICS-profiles are much diminished as compared to those seen in Figure~\ref{fig:Fig3}e. This feature of $\mathrm{NICS}_{\pi zz}$-scans is possibly due to $\sigma$-contamination (more discussion on this in the following section). In fact, $\mathrm{NICS}_{\pi zz}$-scans of borazine show that these minima do not merge till the ghost atoms are about 2.0\,\AA{} above the molecular plane. At this height, the $\mathrm{NICS}_{\pi zz}$ minima coincide with the $\mathrm{NICS}$-value at the center of the ring (up to $0.2\times10^{-3}$ ppm).  Hence, since even $\mathrm{NICS}_{\pi zz}$-scans for borazine show $\sigma$-contamination below 2.0\,\AA{}, we have used 2.0\,\AA{} as the probe height for both hBN and graphene flakes for the sake of uniformity (instead of 1.7\,\AA{} for graphene flakes).  Also, at 2.0\,\AA{}, the minima in NICS(2)$_{zz}$ X-scan for borazine are very shallow, being only  3.8\% lower than the NICS(2)$_{zz}$ value at the center of the ring. In other words, at 2.0\,\AA{}, we are high enough that the $\sigma$-effects are negligible [see Figure S1b], while the $\pi$-contributions have not started decaying appreciably.  We note that NICS-indices determined at a probe-height of 1.7\,\AA{} give qualitatively similar results [Figure S2 in SI].

Along with the aforementioned determination of the optimal height for the BQ-probes, a further comparison of the $\mathrm{NICS}_{zz}$ and $\mathrm{NICS}_{\pi zz}$ scan profiles for borazine and benzene highlight the differences in their aromaticities. The removal of the contribution from the $\sigma$ electrons at the lower heights within the SOM approach has a much greater effect on the NICS-values for borazine than those for benzene. For example, the minima at $r$=1.0\,\AA{} in the $\mathrm{NICS}_{\pi zz}$ curve (see Figure~\ref{fig:Fig3}f) of borazine are 69.4\% shallower than the corresponding minima in the $\mathrm{NICS}_{zz}$ curve in Figure~\ref{fig:Fig3}e. On the other hand, the decrease in the depth of the minimum at $r$=1.0\,\AA{} for benzene's $\mathrm{NICS}_{\pi zz}$ curve is merely 11.5\% as compared to its  $\mathrm{NICS}_{zz}$ curve.  It is natural to suspect that the following two effects (either operating together or in isolation) may be playing a role: (i) there is an overall greater contribution by $\sigma$ electrons to the NICS-values for borazine than for benzene and so, elimination of the $\sigma$ electrons contribution reduces NICS values by a greater extent, and/or (ii) the $\pi$-contribution to NICS itself may be much smaller for borazine than for benzene. To see which of the two effects plays the  dominant role, we plotted the $\mathrm{NICS}_{\sigma zz}$ curves for the fictitious SOM models of borazine and benzene (see Figures S1c and d, and corresponding discussion in Section S1 of SI). This analysis shows that it is the latter effect that is responsible for the lower depths of the minima in SOM's $\mathrm{NICS}_{\pi zz}$ curves of borazine. This conclusion is in agreement with those reached by Carion \textit{et al.}~\cite{carion2010} and B\'{a}ez-Grez \textit{et al.}~\cite{baez2018} using isochemical shielding surfaces.

\subsubsection{Single-point $\textrm{NICS}_{zz}$ and $\textrm{NICS}_{zz}$ XY-scans at a height of 2.0\,\AA {}}
\begin{figure*}[t]
    \centering
    \includegraphics[width=0.99 \linewidth]{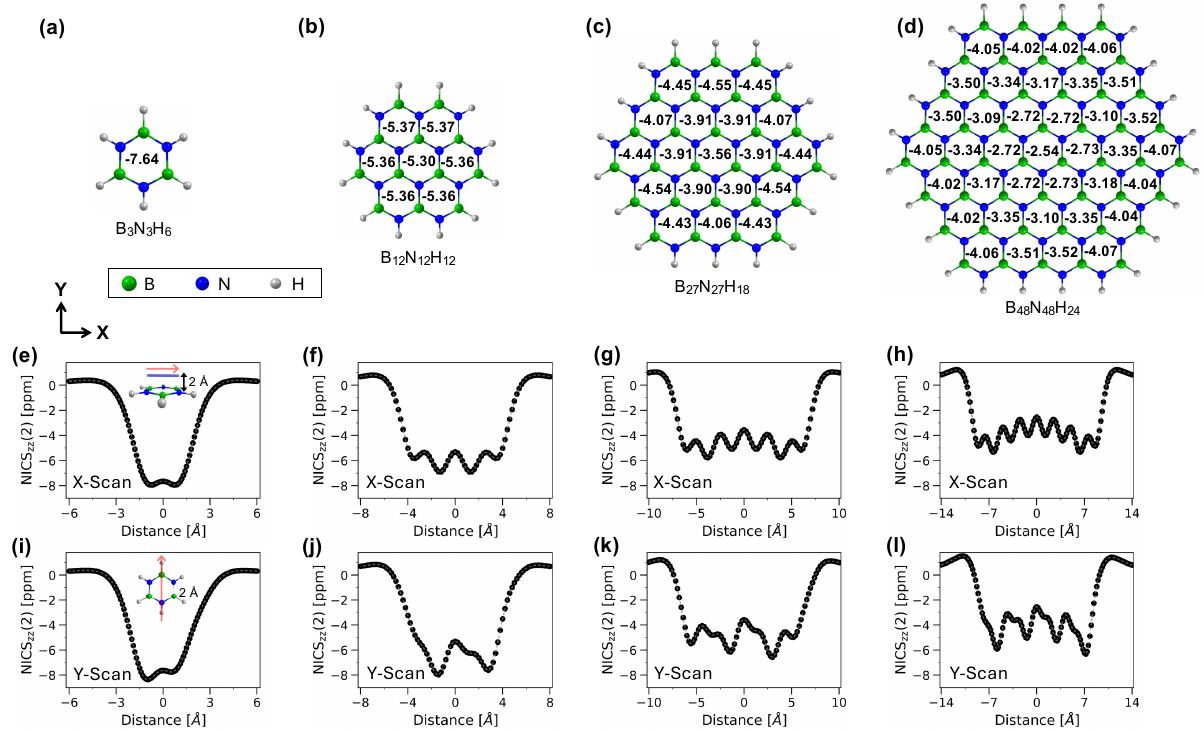}
    \caption{$\textrm{NICS}_{zz}$ values at 2.0\,\AA{} above the ring centers for (a) B$_{3}$N$_{3}$H$_{6}$, (b) B$_{12}$N$_{12}$H$_{12}$, (c) B$_{27}$N$_{27}$H$_{18}$, and  (d) B$_{48}$N$_{48}$H$_{24}$. $\textrm{NICS}_{zz}$ scans along: (e, f, g, h)  X-direction and (i, j, k, l) Y-direction for the hBN flakes of different sizes. The insets in (e) and (i) show the scan directions for borazine. }
    \label{fig:Fig4}
\end{figure*}

Single-point NICS indices have proven to be useful in determining the aromaticity patterns in polycyclic aromatic hydrocarbons. In order to do so for the hBN and graphene flakes, we placed the BQ probes at the center of each ring at a height of 2.0\,\AA {} above the molecular plane of the flakes and performed single-point NICS calculations.  The single-point $\textrm{NICS}_{zz}(2)$ values for hBN flakes are displayed in Figure~\ref{fig:Fig4}a--d.  The negative values of $\textrm{NICS}_{zz}(2)$-indices implies a diatropic response to the applied magnetic field.  These numbers, however, are smaller in magnitude than those obtained for the corresponding graphene flakes (see Figures S3a--d), revealing a weaker aromatic character of hBN.  The single-point NICS analysis shows two trends: (i) we get smaller $\textrm{NICS}_{zz}(2)$ values as the flake size increases, showing size-dependence of NICS values, and (ii) in the case of the polycyclic hBN flakes, the peripheral rings exhibit larger negative NICS-indices as compared to the inner BN-rings, gradually increasing in magnitude from the center to the outer rings. This indicates the presence of stronger diatropic field around the edges. The overall aromaticity pattern displayed by the hBN flakes is different from the aromaticity patterns seen in the graphene flakes. For C$_{24}$H$_{12}$, the $\textrm{NICS}_{zz}(2)$-index of the central ring is smaller than that of the surrounding rings, while for C$_{54}$H$_{18}$, the central ring has the highest $\textrm{NICS}_{zz}(2)$-index, with the other rings showing an identifiable alternating pattern of low and high indices, as was also seen in Zdetsis's work~\cite{Zdetsis_graphene2020}. The differences in trends between hBN and graphene highlight the critical role of chemical composition and bonding character in governing aromaticity. 

In order to spatially resolve the effects of local and global (induced) ring currents~\cite{Stanger_2014_nicsscan}, we also performed the  NICS XY-scans for hBN and graphene flakes by placing a string of BQ-probes along X- and Y-trajectories, and at a height of 2.0\,\AA. 
The results of the X- and Y-scans of hBN flakes are plotted in Figures~\ref{fig:Fig4}e--h and~\ref{fig:Fig4}i--l, respectively. The corresponding plots for the graphene flakes are provided in the Supplementary Figures S3e--h and S3i--l.   For both hBN and graphene flakes of all sizes, we obtained negative $\textrm{NICS}_{zz}(2)$ values within the circumference of the flakes, with the indices becoming positive outside the flakes, before approaching zero as we move further away.

\begin{figure}[ht]
    \centering
    \includegraphics[width=0.82 \linewidth]{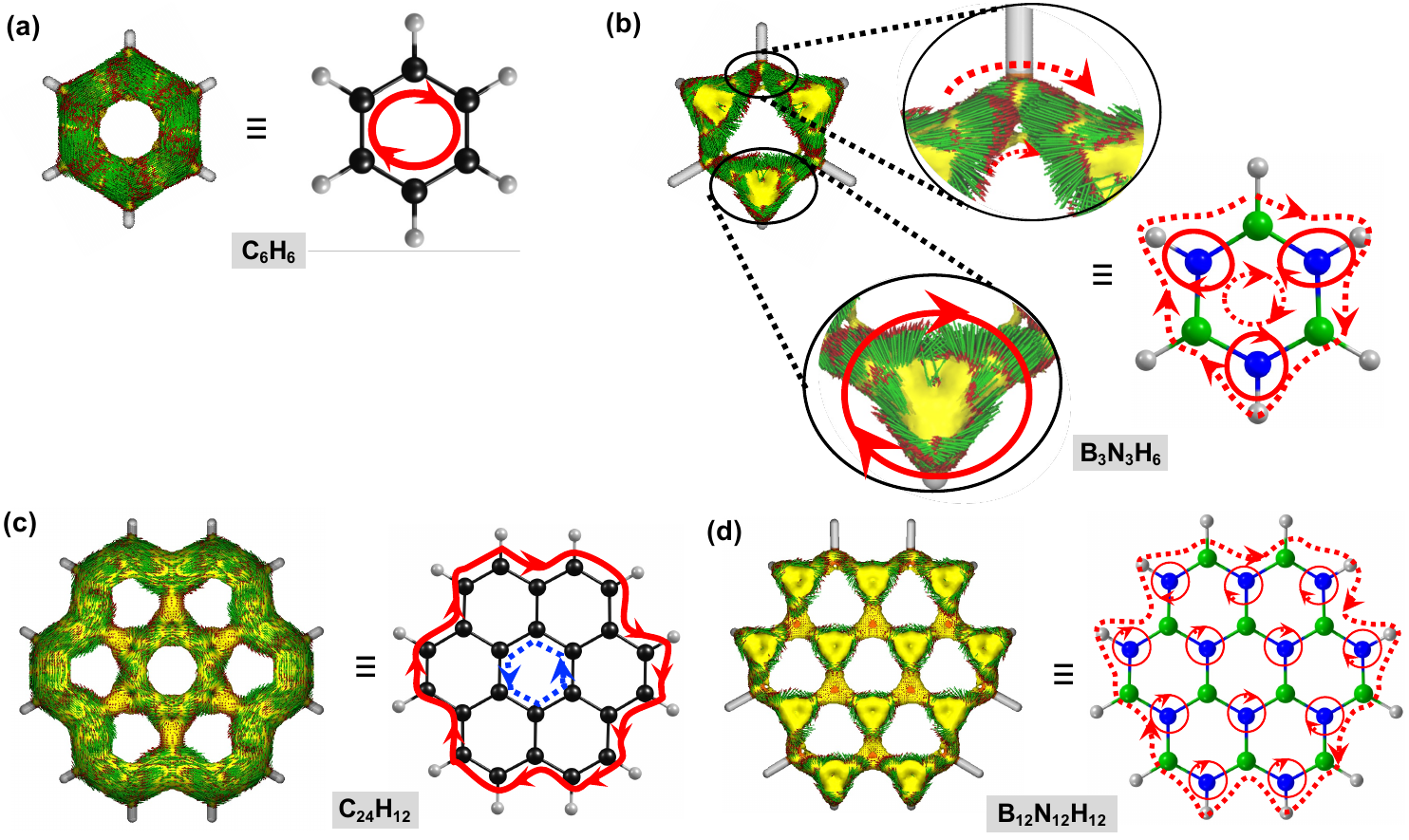}
    \caption{(a) ACID isosurface plot for benzene, showing the $\pi$-contribution. Current density vectors (green arrows) plotted onto the ACID isosurface (drawn in yellow) show a strong diatropic ring current from $\pi$-delocalization. Also drawn is a schematic illustration of the benzenic ring current (solid red line).  (b) ACID plot for borazine showing the $\pi$-contribution. A close-up of different regions shows a weak global diatropic circulation (in dashed red line), along with stronger local ring current (in solid red) around the nitrogen atoms. The schematic diagram shows global and local rings currents to highlight the differences in magnetic response for benzene and borazine. (c) ACID and schematic plots for C$_{24}$H$_{12}$, showing a strong global ring current (diatropic), along with a weak local paratropic current (in blue) around the central ring. (d) ACID and schematic plots for B$_{12}$N$_{12}$H$_{12}$, showing a global, diatropic ring current in the outer rim, along with local diatropic currents around nitrogen atoms. All ACID plots are for the isovalue of 0.03.}
    \label{fig:Fig5}
\end{figure}
A closer inspection of the X- and Y-scans in Figures~\ref{fig:Fig4}e--h and~\ref{fig:Fig4}i--l, respectively, reveals different patterns that are formed by the local maxima and minima in the $\textrm{NICS}_{zz}(2)$-profiles for the hBN flakes. These XY-scan patterns also differ from those formed by graphene flakes [Figures S3e--h and S3i--l].  In order to develop an overall picture from these X- and Y-scans, it is instructive to understand their profiles for benzene and borazine.  Supplementary Figures~S3e and S3i show the NICS X-scan and Y-scan, respectively, for benzene. Both scans have a well-defined minimum at the ring center, indicating that the induced field is strongest there, and that there is a diatropic ring current.

Our conclusions based on the magnetic criterion are supported by the anisotropy of the current (induced) density (ACID) method~\cite{Herges2005_ACID} (see Methods section for details).  Figure~\ref{fig:Fig5}a is the ACID plot for benzene (isovalue 0.03), showing the six-electron diatropic (clockwise) current.  In the case of borazine, the X-scan in Figure~\ref{fig:Fig4}e shows two shallow minima above the bonds, which are possibly due to the residual effects of $\sigma$-electrons. Ignoring the $\sigma$-contamination effects, the X-scan in Figure~\ref{fig:Fig4}e indicates the presence of a diatropic ring current in borazine. Although much weaker, borazine's ring current is benzenic in character.  The Y-scan of borazine in Figure~\ref{fig:Fig4}i shows two unequal minima, with the deeper minimum located above the N-atom and the shallower minimum above the B-atoms of borazine. The deeper minimum at the N atom indicates a stronger induced field at its location as compared to the center and at the B-atom.  It should be pointed out that the change in the $\textrm{NICS}_{zz}(2)$ values at the center and at the local minimum above the B atom is merely 0.12 ppm, and one can treat it as if the induced field (for all purposes) is constant between the center of the ring and the B atom, while it is larger at the N atoms.  In order to understand the asymmetry in the profile of the Y-scan, we considered the ACID plot for borazine, which is shown in Figure~\ref{fig:Fig5}b. The ACID plot shows the contribution of the $\pi$-electrons, with the enlarged portions of the plot showing the presence of a weak global diatropic ring current, along with diatropic currents around the nitrogen atoms. These results are consistent with an earlier report of local and global currents in borazine using magnetically induced current density~\cite{baez2022}. In Figure~\ref{fig:Fig5}b, we have also shown a schematic illustration of the weak global ring (dashed red line) and the strong local (in solid red) circulations. This diagram illustrates why, in spite of some cancellation between the diatropic ring and local currents at the N-atom sites, the field at N-atoms is stronger than that around the B atoms.

The aforementioned arguments can be applied to understand the NICS-profiles of the larger hBN and graphene flakes as well, with their heterogeneity along the X- and Y-directions coming from different chemical environments along these directions.  For the sake of completeness, Figures~\ref{fig:Fig5}c and d show the ACID plots for the larger graphene (C$_{24}$H$_{12}$) and hBN flakes (B$_{12}$N$_{12}$H$_{12}$).  In the case of C$_{24}$H$_{12}$ in Figure~\ref{fig:Fig5}c, we observe a strong diatropic circulation along the outer rim of the molecule.  It does not show benzenic currents around each ring since any current along an inner C--C bond of a ring is cancelled by another current flowing in opposite direction along the same bond shared by the adjoining ring.  A much smaller paratropic current can also be seen in the central ring. These features are also captured by the X- and Y-scans for  C$_{24}$H$_{12}$ [Supplementary Figures S3f and j], and were reported earlier~\cite{Stanger_2014_nicsscan}. In the case of the hBN flake (B$_{12}$N$_{12}$H$_{12}$), once again the ACID plot does not display weak benzenic currents around each ring due to the cancellation of currents along the inner B--N bonds that are shared by two rings, just as in the case of C$_{24}$H$_{12}$.  However, in spite of this similarity, the $\pi$ circulation in the hBN flake shows a distinct pattern when compared to the corresponding graphene flake. The ACID plot for B$_{12}$N$_{12}$H$_{12}$ shows a much weaker global current (diatropic) along the outer rim of the flake, along with the local diatropic currents around the N atoms. 

Overall, the XY-scans and the ACID plots reveal a weaker diatropic response in hBN flakes as compared to the graphene flakes. In order to understand the weaker aromaticity of hBN, we further investigated the nature of frontier orbitals and how the symmetry properties of $\pi$ orbitals might be playing a role.

\subsubsection{Frontier molecular orbitals and symmetry-allowed transitions}

\begin{figure*}[ht]
    \centering
    \includegraphics[width=0.9 \linewidth]{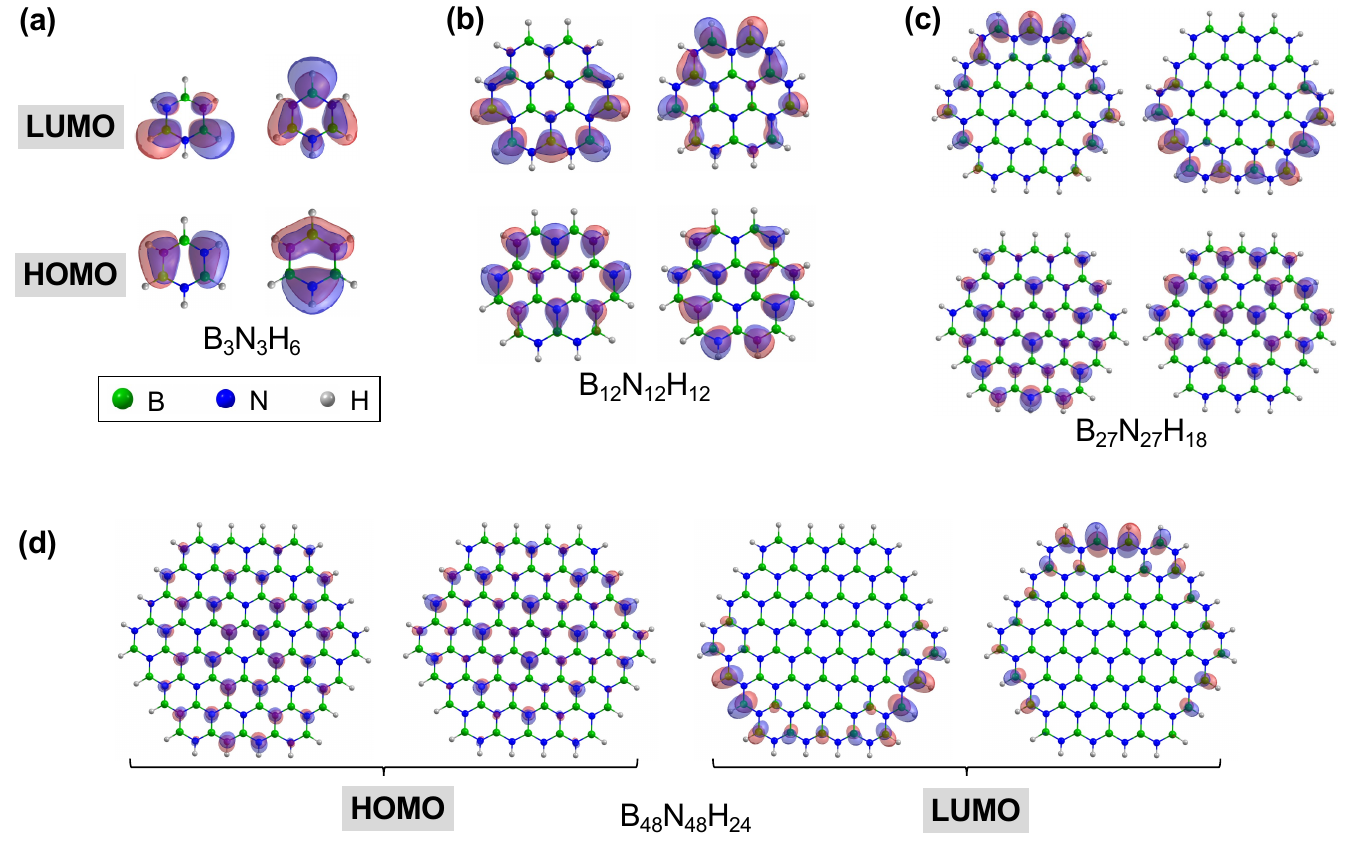}
    \caption{The frontier molecular orbitals (MO) -- highest occupied MOs (HOMOs) and lowest unoccupied MOs (LUMOs) of: (a) B$_{3}$N$_{3}$H$_{6}$, (b) B$_{12}$N$_{12}$H$_{12}$, (c) B$_{27}$N$_{27}$H$_{18}$, and (d) B$_{48}$N$_{48}$H$_{24}$. Both the HOMO and LUMO for all of the hBN flakes belong to the $e''$-representation.}
    \label{fig:Fig6}
\end{figure*}

 Figures~\ref{fig:Fig6}a--d and the Supplementary Figures S4a--d show the frontier molecular orbitals (MOs) for hBN and graphene flakes, respectively. For all hBN flakes ($D_{3h}$ symmetry), both the highest occupied MOs (HOMOs) and lowest unoccupied MOs (LUMOs) belong to the $e''$ representation. On the other hand, the symmetry representations of HOMOs and LUMOs for the graphene flakes ($D_{6h}$ symmetry)  switch between the $e_{1g}$ and $e_{2u}$ representations. Also note that the HOMO and LUMO of borazine (Figure~\ref{fig:Fig6}a) show an increased electron density on the N-atom due to its higher electronegativity. On the other hand, the frontier orbitals of benzene show a more uniform distribution of the charge cloud between two consecutive C-atoms. For both benzene and borazine, the frontier orbitals are delocalized over their respective molecule, which is expected for aromatic molecules. This is also the case for the other smaller flakes.  On the other hand, for the larger hBN and graphene flakes, the frontier orbitals (especially, the LUMOs of B$_{48}$N$_{48}$H$_{24}$ in Figure~\ref{fig:Fig6}d and C$_{96}$H$_{24}$ in Figure~S3d) show a greater localization over the rim/edges of the flakes.  In spite of these differences, the frontier orbitals show an overall delocalization and hence, aromaticity of the hBN and graphene flakes.  
 
 Although the differences between the charge densities corresponding to frontier orbitals of hBN and graphene flakes may hint at their different aromatic character, it is the relationship between orbital symmetries and allowed orbital transitions under the influence of external magnetic field~\cite{Fowler2001,Fowler2004,Fowler2007} that provides a more direct rationalization of the smaller values of induced ring current (and hence, lower NICS values) for the hBN flakes.  In what follows, we use the symmetry arguments to explain these differences between borazine and benzene. The same symmetry arguments can be applied to the larger graphene and hBN flakes, as the relevant transition amplitudes involve orbitals having the same symmetries in the larger flakes as those involved in benzene and borazine.

\begin{figure*}[ht]
    \centering
    \includegraphics[width=0.95 \linewidth]{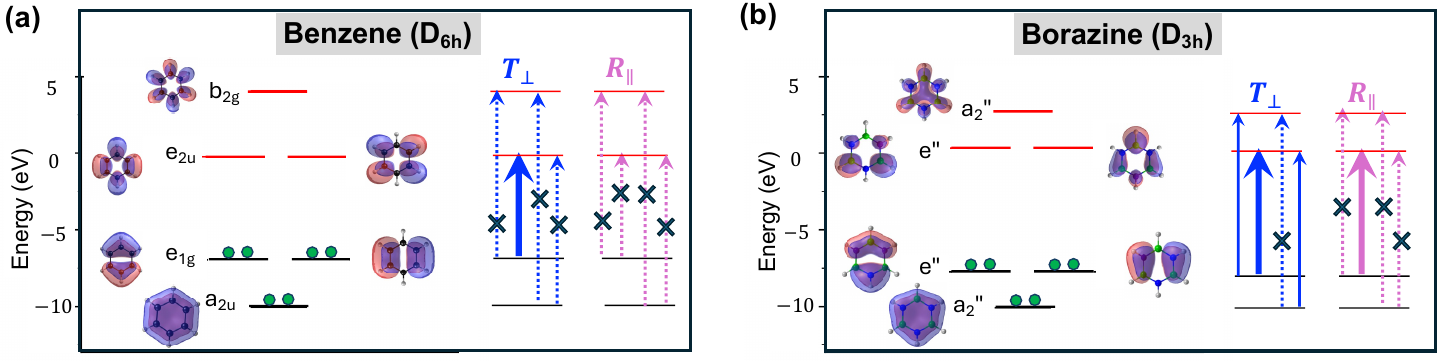}
    \caption{Energy-level diagram for (a) benzene and (b) borazine, showing virtual excitations between the filled (denoted by filled green circles) and empty $\pi$-orbitals that result in either diatropic (translational transitions) or paratropic (rotational transitions) ring currents, or both.  The solid blue (pink) arrows are used to indicate symmetry-allowed translational (rotational) transitions. The dashed arrows (with a cross) show transitions that are symmetry-forbidden.  Different arrow thicknesses for the allowed transitions indicate different relative weights, with the weight of a transition amplitude being inversely proportional to the energy difference between the involved orbitals. The relevant operators involved in transitions are two right-angle translations (labelled as $T_{\perp}$) and the rotation about the field direction ($R_{\parallel}$). }
    \label{fig:Fig7}
\end{figure*}

Fowler and coauthors~\cite{Fowler2001,Fowler2004,Fowler2007} showed that the ring currents are a consequence of transitions between occupied and unoccupied orbitals upon application of an external magnetic field. Assuming that the applied magnetic field is perpendicular to the molecular plane [(x,y) plane], the operators that give rise to the diatropic ring currents are the two translations within the molecular plane [$T_{\perp}=(T_{x}, T_{y})$], and the operator involved in the paratropic transitions is the rotation around the field direction ($R_{\parallel} \equiv R_{z}$).  Then a transition between unoccupied and occupied $\pi$ orbitals, $|\psi_{i}\rangle \longrightarrow |\psi_{f}\rangle$, will be allowed and give rise to diatropic current if the product $\Gamma(\psi_{i}) \otimes  \Gamma(T_{\perp}) \otimes  \Gamma(\psi_{f})$ contains a totally symmetric representation. Here, $\Gamma(\psi_{i})$, $\Gamma(\psi_{f})$, and $\Gamma(T_{\perp})$ are the representations of the occupied orbital, unoccupied orbital, and the translation operator, respectively. Similarly a transition will result in a paratropic response if the product $\Gamma(\psi_{i}) \otimes \Gamma(R_{\parallel}) \otimes \Gamma(\psi_{f})$ contains a totally symmetric representation.  Hence, depending on the symmetry of the pair of involved $\pi$-orbitals, an allowed transition may give rise to a diatropic or a paratropic ring current, or even both.  Fowler and coauthors also showed that the strength of the contribution is inversely proportional to the energy difference between the involved orbitals, and that the magnitude of contribution to the current is also determined by the spatial overlap between the orbitals.  We first apply these considerations to benzene ($D_{6h}$ symmetry).  Within $D_{6h}$, the two perpendicular translations and the rotation around the z-direction transform as the $e_{1u}$ and $a_{2g}$ representations, respectively.  We find that the only allowed translational transition is between HOMO ($e_{1g}$) and LUMO ($e_{2u}$), resulting in a diatropic current (since $e_{1g} \otimes e_{1u} \otimes e_{2u}$ contains $a_{1g}$), while no rotational transition is allowed. The energy level diagram in Figure~\ref{fig:Fig7}(a) summarizes this analysis. In contrast to benzene, the lower symmetry of borazine ($D_{3h}$) results in a more complicated picture, which is shown in Figure~\ref{fig:Fig7}(b). In borazine, both HOMO and LUMO belong to the $e''$-representation. Also, within the $D_{3h}$-symmetry, the perpendicular translations and the rotation around the applied field transform as $e'$ and $a'_{2}$, respectively.  A transition between HOMO and LUMO is active under both translation and rotation, implying that both diatropic and paratropic currents are generated. These should effectively cancel each other out and one would expect that borazine should be non-aromatic.  However, in addition to the virtual transitions between HOMO and LUMO, we find that two additional translational transitions are allowed between the $\pi$ orbitals as shown in Figure~\ref{fig:Fig7}(b). Since these transitions have smaller relative weight due to a greater energy difference between the involved $\pi$-orbitals, these transitions result in a weaker diatropic current, and hence, smaller NICS-values for borazine.

\subsection{Energetic Criterion of Aromaticity: Aromatic Stabilization Energy (ASE)}

 \begin{figure*}[t]
    \centering
    \includegraphics[width=0.8 \linewidth]{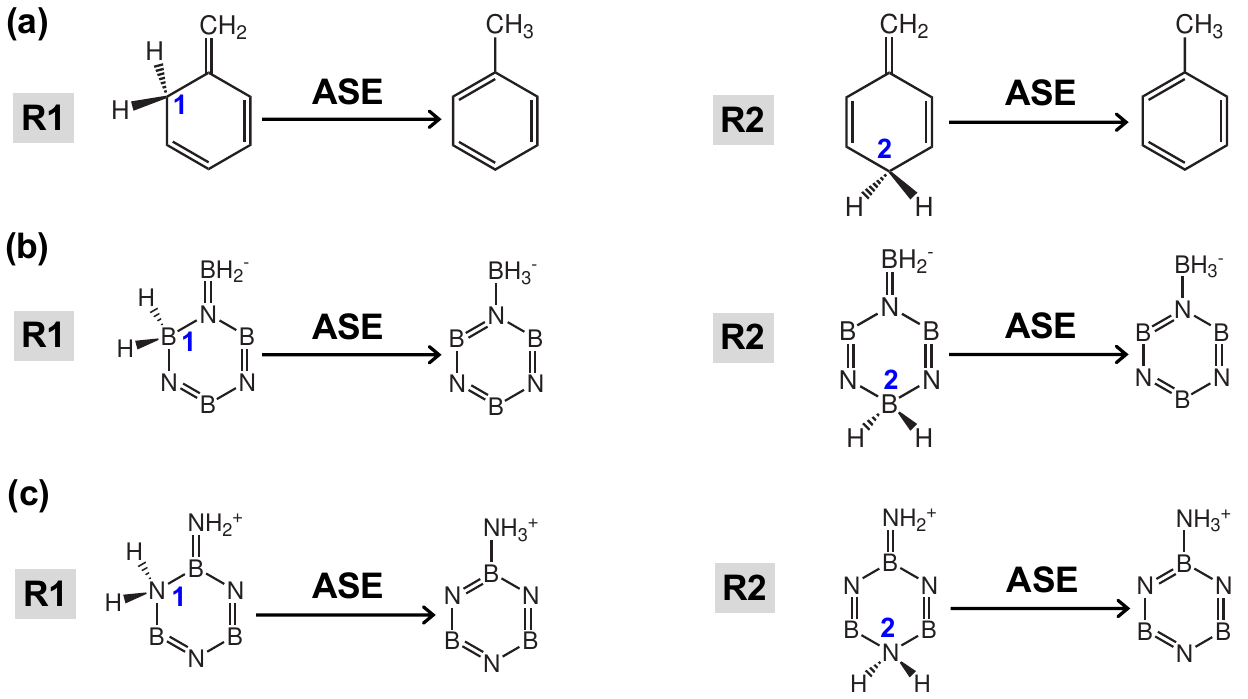}
    \caption{Energetic determination of aromaticity of graphene and hBN flakes using isomerization method: (a) The isomerization reactions (labelled as R1 and R2) used to evaluate the aromatic stabilization energy (ASE) for benzene. Toluene (\ce{$\mathrm{C_{6}H_{5}}$-$\mathrm{CH_{3}}$}) is used as a stand-in for benzene.  Ortho-isotoluene and para-isotoluene are used as the reference non-aromatic reactants in reactions R1 and R2, respectively. Reactions used to evaluate ASE for borazine, with the (b) borane anion-substituted (\ce{$\mathrm{B_{3}N_{3}H_{5}}$-$\mathrm{BH_{3}^{-}}$}) and (c) protonated amino group-substituted (\ce{$\mathrm{B_{3}N_{3}H_{5}}$-$\mathrm{NH_{3}^{+}}$}) structures as stand-ins for borazine.}
    \label{fig:Fig8}
\end{figure*}
 In order to support our conclusions based on the magnetic criterion for aromaticity determination, we also calculated the aromatic stabilization energy (ASE) values. ASE is defined as the difference in the total energies of an aromatic molecule and its non-aromatic counterpart~\cite{Pauling_ASE_1933}.  The non-aromatic reference structure can be chosen in a number of ways, yielding a wide range of ASE values. As a result, even for the canonical aromatic molecule -- benzene -- the calculated ASE values diverge from each other by more than 50 kcal/mol (see the review article by Cyra\'{n}ski~\cite{Cryanski_review} and the references therein).  This variation in ASE values of benzene is due to a number of factors, including strain, different types of bonds, and presence of heteroatoms in the non-aromatic reference structures.  These drawbacks are removed within the isomerization method proposed by Schleyer \textit{et al.}~\cite{schleyer2002_ISE}. An added advantage of using the isomerization method within our work was that it could be easily generalized to the larger hBN and graphene flakes.  Within this method,  ASE is calculated as a difference in the energy of the product, which is a proxy to the aromatic molecule of interest, and a reactant, which is a non-aromatic reference isomer of the proxy. Figure~\ref{fig:Fig8}a shows two such isomerization reactions, labelled as R1 and R2, for estimating the ASE of benzene. In both of these reactions, toluene (\ce{$\mathrm{C_{6}H_{5}}$-$\mathrm{CH_{3}}$}) is employed as an aromatic stand-in for benzene. This is done because toluene, unlike benzene, allows one to create its non-aromatic isomers with an identical number and type of C-C bonds in their structures. The non-aromatic isomers can be created by shifting a hydrogen from the methyl group to the ortho position (labelled as ``1'') or the para  position (labelled as ``2''), resulting in ortho-isotoluene in the former case and para-isotoluene in the latter case.  The ASE is then obtained from the difference between the total energies of the aromatic product (\ce{$\mathrm{C_{6}H_{5}}$-$\mathrm{CH_{3}}$}) and one of the non-aromatic reactants. 
 In switching to toluene to obtain the ASE for benzene, the assumption is that toluene and benzene have nearly-identical aromatic properties~\cite{schleyer2002_ISE}. We show this explicitly in the next section.  
These reactions were generalized to larger flakes of graphene by creating methyl-derivatives of these conjugated aromatic structures, and comparing their energies with those of their respective non-aromatic isomers [see Figure S5 in Supporting Information (SI)].  Since hBN is a binary compound, we considered two possible groups (see Figures~\ref{fig:Fig8}b and c), as done previously for borazine~\cite{baez2022}. In these figures, the borane anion-substituted structure (\ce{$\mathrm{B_{3}N_{3}H_{5}}$-$\mathrm{BH_{3}^{-}}$}) and protonated amino group-substituted structure (\ce{$\mathrm{B_{3}N_{3}H_{5}}$-$\mathrm{NH_{3}^{+}}$}) are stand-ins for borazine, allowing one to create non-aromatic counterparts that keep the total number and type of B-N bonds unchanged. These reactions were then generalized to obtain ASEs for the larger hBN flakes.

In order to make an aromaticity determination based on energetic criterion, we first established that the products in Figures~\ref{fig:Fig8}a--c are good stand-ins for benzene and borazine. Toluene (\ce{$\mathrm{C_{6}H_{5}}$-$\mathrm{CH_{3}}$}) has a $\textrm{NICS}_{zz}(2)$ index of $-16.85$\,ppm (see Supplementary Figure S6a), which is comparable to the $\textrm{NICS}_{zz}(2)$ index of benzene ($-17.41$\,ppm from Supplementary Figure S3a). 
This shows that toluene is a good proxy for benzene. In addition, toluene's isomers, ortho- and para-isotoluene, have $\textrm{NICS}_{zz}(2)$ indices of $-2.76$ and $-4.43$\,ppm, respectively, which implies that the $\pi$-delocalization is significantly diminished in the two reactants (see Supplementary Figure S6a).  Hence, ortho- and para-isotoluene molecules are good non-aromatic references. The two reactions -- R1 and R2 -- yield ASE values of -34.18 and -30.31\,Kcal/mol, respectively, for benzene. These results are consistent with values reported in earlier works~\cite{schleyer2002_ISE}, confirming the aromaticity of benzene.  Note that as shown by Schleyer and P\"{u}hlhofer~\cite{schleyer2002_ISE}, the difference in ASE values calculated for R1 and R2 is due to anti-syn diene mismatch.  

In the case of \ce{$\mathrm{B_{3}N_{3}H_{5}}$-$\mathrm{BH_{3}^{-}}$} and \ce{$\mathrm{B_{3}N_{3}H_{5}}$-$\mathrm{NH_{3}^{+}}$}, which are used as two proxies for borazine, we obtain $\textrm{NICS}_{zz}(2)$ indices of $-7.78$, and $-7.60$\,ppm, respectively. These values are comparable to the $\textrm{NICS}_{zz}(2)$ index of $-7.64$\,ppm for borazine [see Figure~\ref{fig:Fig4}a], implying that borazine and its two stand-ins have nearly identical aromatic properties. 
The two reactants containing the \ce{=$\mathrm{BH_{2}^{-}}$} group in reactions R1 and R2  have $\textrm{NICS}_{zz}(2)$ indices of $-2.25$  and $-3.23$\,ppm, respectively (see Figure~\ref{fig:Fig8}b and Supplementary Figure S6b). On the other hand, the two isomers containing the \ce{=$\mathrm{NH_{2}^{+}}$} group have $\textrm{NICS}_{zz}(2)$ indices of $-4.50$  and $-4.90$\,ppm, respectively (see Supplementary Figure S6c). This implies that the \ce{=$\mathrm{BH_{2}^{-}}$}-containing isomers are better non-aromatic reference molecules as compared to \ce{=$\mathrm{NH_{2}^{+}}$}-containing isomers, although the latter are less aromatic than borazine itself. Hence, the ASE values for borazine estimated from the isodesmic reactions in Figure~\ref{fig:Fig3}b are expected to be larger than those obtained from the reactions in Figure~\ref{fig:Fig8}c.  We find that this is indeed the case. The ASE values for reactions R1 and R2 in Figure~\ref{fig:Fig8}b are -12.55 and -10.82\,Kcal/mol, respectively, while the corresponding values for reactions in Figure~\ref{fig:Fig8}c are -6.73 and -6.79\,Kcal/mol, respectively. The latter can be considered as the lower limit values for aromaticity of hBN's flakes. Our calculated ASE values for borazine are about 36-37\% of that for benzene when considering reactions in Figure~\ref{fig:Fig8}b and about 20-22\% of that for benzene when considering reactions in Figure~\ref{fig:Fig8}c. These values are in agreement with those reported by B\'{a}ez-Grez \textit{et al.}~\cite{ baez2022}, and show that borazine is less aromatic than benzene. 

The reactions in Figure~\ref{fig:Fig8}a--c are  generalized to larger flakes (see Supplementary Figures~S5, S7--S9 and the Table~S1).  Even with the simple generalization of the isomerization method, which uses only a  single functional group substitution, our results tabulated in the Supplementary Table~S1 support our conclusion that  hBN is aromatic,  even if it is less so as compared to graphene. Note that although consideration of different numbers of functional group substitutions for larger flakes, along with all of their stable, low-energy conformations, or other effects that may influence the ASE values (such as hydrogen-hydrogen repulsion and strain) is beyond the scope of this work, additional studies including these factors may help to further highlight the differences in hBN's and graphene's aromatic properties.

  \section*{Conclusions}

Since hBN is a technologically important material, a better understanding of its fundamental properties is important and relevant to its different applications. In experiments, hBN is routinely used as an encapsulation layer/substrate for other low-dimensional materials. It is also being explored as a host of defect-based quantum emitters for quantum sensing and computing applications. All of these applications rely on it being a bench-stable material. 
However, till now, it was not known if hBN's aromaticity may be contributing to its chemical stability. In this work, we used  magnetic criterion, supported by group-theoretic considerations, as well as the energetic criterion, to investigate hBN's aromaticity.  hBN displays negative $\textrm{NICS(2)}_{zz}$ indices in its rings, satisfying the magnetic criterion for aromaticity. It is also lower in energy than its non-aromatic counterpart(s). Hence, different indicators of aromaticity are mutually consistent and show that hBN is indeed aromatic, even though it is less so as compared to graphene. Finally, our symmetry-based argument explained why hBN's aromatic properties are distinct from those of graphene.

  \section*{Methods}

In this density functional theory (DFT)-based study, we used hexagonal flakes of different sizes (chemical formula: B$_{3n^{2}}$N$_{3n^{2}}$H$_{6n}$, $n=1, 2, 3,...$). The flakes are shown in Figure~\ref{fig:Fig1}. The same sequence of flakes was also created for graphene (chemical formula: C$_{6n^{2}}$H$_{6n}$). All DFT calculations were performed at the B3LYP/6-311G(d,p)~\cite{Becke1988-BLYP,Lee1988-BLYP} level of theory using Gaussian 16 program (Revision C.01)~\cite{g16}.  We also ensured that the fully-optimized flakes are dynamically stable and do not have imaginary vibrational modes.

 For the magnetic criterion, we employed Schleyer's nucleus independent chemical shift (NICS) method~\cite{Schleyer_1996_NICS,Schleyer_1997_NICS1} using the standard gauge invariant atomic orbital (GIAO) approximation as implemented in the Gaussian 16 package (Rev C.01). The chemical shift calculations for all of the flakes were performed using B3LYP exchange correlation functional and the 6-311G(d,p) basis set.  Both the single-point  and multi-point NICS approaches were used in order to create a fuller picture of hBN's aromaticity.  The conclusions based on the magnetic criterion are supported by the group theoretic considerations, as well as, the anisotropy of the current (induced) density (ACID)~\cite{Herges2005_ACID}. The ACID plots were generated using the continuous set of gauge transformations (CSGT) approach, providing a visual representation of the induced ring currents in the molecular models for hBN and graphene.

Aromatic stabilization energy (ASE), which is defined as the difference in the total energies of an aromatic molecule and its non-aromatic counterpart~\cite{Pauling_ASE_1933}, was also calculated. We used the isomerization method proposed by Schleyer \textit{et al.}~\cite{schleyer2002_ISE} to obtain the ASE values.  Within this method,  ASE is calculated as a difference in the energy of the product, which is a proxy to the aromatic molecule of interest, and a reactant, which is a non-aromatic reference isomer of the proxy (see Figures~\ref{fig:Fig8}a--c).  These reactions were then generalized to obtain ASEs for the larger hBN flakes.

\vspace{1.0cm}

\section{Associated Content}
\noindent \textbf{Data Availability Statement}\\
The crystal structures data that support the findings of this study are available
from the corresponding author upon reasonable request.

\noindent \textbf{Supporting Information}\\
\noindent The Supporting Information is available free of charge.

\noindent Supplementary material includes information regarding the generalization of isomerization method for determining aromatic stabilization energies of polycyclic flakes and results from this analysis, along with additional details to explain the magnetic response of hBN as compared to graphene.

\section{Author Information}
\noindent \textbf{Corresponding Author}\\

\noindent \textbf{Pratibha Dev} -- Laboratory for Physical Sciences, College Park, Maryland 20740, United States;  orcid.org/0000-0002-6884-6737;
Email: pdev@lps.umd.edu\\

\noindent \textbf{Authors}\\
\noindent \textbf{Suryakanti Debata} -- Laboratory for Physical Sciences, College Park, Maryland 20740, United States; orcid.org/0000-0003-4983-3065; Email: sdebata@lps.umd.edu\\
\noindent \textbf{Sai Krishna Narayanan} --  Department of Physics, University of Maryland, College Park, Maryland 20742, United States; orcid.org/0000-0003-1420-4488; Email: saik98@umd.edu\\

\noindent \textbf{Notes}\\
The authors declare no competing financial interest.

\begin{acknowledgement}
\noindent \textbf{Acknowledgements:} 
This work used the Bridges2 clusters at PSC through allocation PHY180014 from the Advanced Cyberinfrastructure Coordination Ecosystem: Services \& Support (ACCESS) program, which is supported by National Science Foundation grants No. 2138259, No. 2138286, No. 2138307, No. 2137603, and No. 2138296.  The authors also thank Prof. Rainer Herges for providing the ACID software.

\end{acknowledgement}

\normalem  


\providecommand{\latin}[1]{#1}
\makeatletter
\providecommand{\doi}
  {\begingroup\let\do\@makeother\dospecials
  \catcode`\{=1 \catcode`\}=2 \doi@aux}
\providecommand{\doi@aux}[1]{\endgroup\texttt{#1}}
\makeatother
\providecommand*\mcitethebibliography{\thebibliography}
\csname @ifundefined\endcsname{endmcitethebibliography}
  {\let\endmcitethebibliography\endthebibliography}{}

%

\end{document}